# Study on the formation of $MgB_2$ phase

Qing-rong Feng[1*], Chinping Chen [1], Jun Xu[1,2],
Ling-wen Kong[1], Xiu Chen[1], Yong-zhong Wang[1], and
Zheng-xiang Gao[1]

Department of Physics and State Key Laboratory of Artificial Microstructure and Mesoscopic Physics, Peking University, Beijing 100871, P.R. China

Electronic Microscopic Laboratory of School of Physics, Peking University, Beijing 100871, P.R. China

**Abstract**

Careful investigations on a series of the $MgB_2$ phases quenched at different sintering temperatures during fabrication process have been carried out, and important information is obtained, by the X-ray diffraction, the SEM images, and the temperature dependent magnetization (M-T) measurements. The particle sizes of the raw magnesium and boron powders are on the order of 0.1 mm and 1 μm, respectively. Three stages in the formation of the polycrystalline $MgB_2$ phase are identified in different sintering temperature ranges according to the corresponding superconducting properties and the crystallization conditions. The $MgB_2$ phase first appears in a microscopic scale at about 530°C according to the M-T measurement. Then, the $MgB_2$ phase forms macroscopically within a narrow temperature range from 653°C to 660°C, while the residual raw materials of the magnesium and boron components persist to the temperature of 700°C. The optimum sintering temperature range for the $MgB_2$ phase to form is determined from 750°C to 900°C.

**Introduction**

Ever since the discovery of the superconductivity with the $MgB_2$[1], intensive investigations have been conducted on the superconducting and the normal state properties[2-3]. These include the crystallographic structure, the thermoelectric power[4,5], the critical field property[6], the transport properties[7-9], the doping effect[10-11], the isotope effect[12], the proton irradiation effect[13], the microwave impedance measurements[14-15], and the specific heat[16-17], *etc*. Many physical quantities and parameters, such as Tc, Jc, $\xi$, $\lambda$, $Hc_1$, $Hc_2$, *etc.*, of this material have been determined by experiments. In the meanwhile, various synthesis and fabrication techniques of the samples in the forms of polycrystalline samples[18], thin films[19-20], wires, and tapes[21-22] are under intensive study as well.

For the synthesis of bulk $MgB_2$, attentions have been focused on the improvements of the sample properties and the optimization of the fabrication processes. Knowledge concerning the phase formation certainly helps in producing samples of better properties. There are two of the techniques commonly applied to prepare the polycrystalline samples of the $MgB_2$ superconductor without involving high pressure conditions in the production process, the one in the vacuum conditions; and the other in the flowing argon atmosphere at the ambient pressure. None of these, however, is effective to produce high-density $MgB_2$ bulk sample (HD-$MgB_2$) by a single sintering process. A technique to produce the HD-$MgB_2$ sample by a double sintering process at high temperature and in ambient pressure has been developed [23]. The density of the sample thus obtained exceeds 2.2 g/cm$^3$ easily. Samples of high density contain less micro-cavity. The "quality" of superconductor, such as the critical current density, the transition sharpness, etc., of the $MgB_2$ could be improved accordingly. The *in-situ* high temperature resistance measurements (HT-RT) during the sintering process both in the vacuum and the flowing argon conditions for the fabrication of the HD-$MgB_2$ superconductor have been reported previously. The formation of the superconducting phase occurs within the temperature range from 645 to 700°C, for the sample prepared in the vacuum conditions of $10^{-5}$ torr[24], and from 537°C to 612°C [25] for that prepared in the argon conditions. The dependence of the formation temperature of the superconducting phase upon various raw Mg powders with different particle sizes has been studied also[26]. For the raw materials using nanometer magnesium powder processed in the vacuum condition, low formation temperature is achieved, from 430°C to 490°C. The smaller particle size of the Mg raw material then results in a lower formation temperature.

In order to provide deeper insight into the phase formation process from the raw material stage to the completion of the $MgB_2$ phase, we have applied various experimental techniques, including XRD analysis, SEM imaging, and M-T measurement, to study a series of samples quenched at various temperatures during the sintering process. Interesting information on the evolution of the $MgB_2$ phase formation is obtained for the samples sintered in the vacuum conditions..

**Sample preparation**

The polycrystalline sample was obtained by the solid-state reaction method using as raw materials the regular off-the-shelf Mg powder with the grain size ≤0.1 mm of 99% in purity and the boron powder with the grain size ≤ 1 μm of 99.99% in purity. The powders were mixed thoroughly with a stoichiometric of Mg : B = 1 : 2, and then pressed by a pressure of 600 MPa into a rectangular $MgB_2$ sample with the dimension of 1.9×0.6×0.20 cm$^3$. The $MgB_2$ embryo sample was then put into a soft iron tube, which was subsequently sealed into a quartz tube. Thirteen sets of the raw materials thus prepared were then sintered in the furnace with a rising rate of the temperature at about 400°C per hour. It was easy to proceed the sintering of the thirteen sample sets at the following temperatures, 450°C, 530°C, 600°C, 620°C, 646°C, 653°C, 660°C, 680°C, 700°C, 750°C, 800°C, 900°C, and 980°C, for 10 minutes respectively and then quenched in the liquid nitrogen at the corresponding temperature by breaking the quartz tube.

The powder XRD patterns were performed by a Philip x' pert X-ray diffractometer on the samples after the heat treatment and on the $MgB_2$ embryo sample before the sintering. The SEM images were taken by the Stara BD325 focus ion beam (FIB) electron microscope and the ZFC magnetization measurement was carried out by a SQUID magnetometer (Quantum Design MPMS) with a background field of 50 Oe.

**Result and discussion**

From the previous report concerning the in situ HT-RT measurements during the temperature rising of the sample fabrication process in the vacuum conditions of $10^{-3}$ Pa [24], a bump appears at about 650°C before a dramatic decrease in the resistance occurs as the temperature increases. This bump is an indication of the appearance of the Mg melting phase. As the temperature goes up higher, the resistance drops at a fast rate with the rising temperature to reach a minimum at about 750°C and then increases gradually with the temperature. The optimum sintering temperature for the superconducting phase formation, as determined by the transition temperature, $T_C$, of the fabricated sample is about 750°C. This is consistent with the temperature at which the minimum resistance occurs in the HT-RT result.

In the present work, we have "frozen" a series of phases for the samples at various stages with different sintering temperatures mentioned in the preceding section. The X-ray diffraction patterns of these frozen phases are shown in Figure 1. The signal intensities in the vertical scale are without definite correlation to each other. The peaks for the B and Mg phases are labeled as 1 and 2 in the spectrum of the pre-sintered sample at the bottom of Fig. 1 while the major peaks for the MgB2 phase are indexed in the spectrum of T = 800 °C. Within the detection resolution of the XRD analysis, the Mg and B phases start disappearing in the sample quenched at 653°C, with traces of the residual components vanishing completely at T > 750°C. In the meanwhile, the characteristic peaks for the $MgB_2$ appear at 660°C indicating that the $MgB_2$ phase forms in a macroscopic scale at this temperature. At the sintering temperature of 980°C, on the other hand, peaks other than those for the $MgB_2$ phase show up. This is a sign for the $MgB_2$ phase transforming into another phase. The above evidences indicate that the formation of the $MgB_2$ phase begin within the narrow temperature range from 653°C to 660°C while the optimum temperature for the phase formation is from 750°C to 900°C, consistent with the result of the HT-RT measurement reported previously. Another important indication of the phase formation, which supports the results of the XRD analysis, is the variation of the sample color by the heat treatment. It changes from taupe brown to black at 653°C as temperature rises from below.

The six SEM images, shown in Fig. 2, are for the MgB$_2$ samples with the heat treatment conditions including the pre-sintered one, the ones sintered at 600°C, 653°C, 660°C, 750°C and 980°C. The magnification power is 50000, corresponding to the resolution level of micron size. Three distinct crystal morphologies are observed for the samples, the ones with the sintered temperature below 653°C, above that, and the one equal to 980°C. For those with the sintered temperature, T ≤ 653°C, the characteristics of the grains are similar to that of the pre-sintered one, showing basically the feature of the magnesium and the boron phases with sharp corners and edges of the micron-size grains. For the samples sintered at the range, 653°C < T < 980°C, the observed grains are much bigger with round corners and edges. This indicates that the MgB$_2$ phase begins to form on the micron-size level at the temperature of 653°C. The photo, taken for the one sintered at 980°C, shows features of small grain other than the previous two types. This reflects the fact as indicated by the XRD analysis that the MgB$_2$ phase undergoes a transformation to another phase and is consistent with the result of the magnetization measurement described in the following that the superconducting Meissner state disappears almost completely for this particular sample.

The Meissner diamagnetizm of the superconducting state is investigated by a DC M-T measurement using MPMS (Quantum Design). The magnitude of the diamagnetizm is an indication of the superconducting content in the sample. This is a more sensitive method to probe the existence of the tiny superconducting phase than the one of HT-RT measurement, since the Meissner state can be detected by the magnetization measurement before the superconducting path forms throughout the sample. In Fig. 3, the difference of the DC magnetization for the Meissner state from the normal state in logarithmic function, $\ln(\Delta M)$, is plotted against the sintering temperature of the sample. A minimum appears roughly at T = 530°C, as indicated by the arrow in the figure, then increases to the plateau starting at the sintering temperature of 700°C. This indicates that the superconducting phase appears in a microscopic scale at the temperature of 530°C. This is lower than the phase formation temperature of 653°C determined by the XRD analysis and beyond the resolution level of the SEM characterization at the magnification power of 50000. The low formation temperature at about 530°C is ascribed to the fact that the small amount of nano-particles existing in the magnesium and the boron raw powders would undergo reaction at a lower temperature. This is consistent with the result from the previous experiment that the magnesium and the boron powders, with fine particles in nanometer scale, starts reaction in the temperature range of 430°C to 490°C.

**Conclusion**

The phase formation process of the MgB$_2$ is investigated by three different techniques, the XRD analysis, the SEM observations, and the M-T measurements. Three stages in the temperature of the heat treatment for the formation of the $MgB_2$ phase are identified according to the evidences obtained by the above techniques. For the early stage, the detection of the Meissner state by the M-T measurement reveals evidence for the existence of the $MgB_2$ superconducting phase at 530°C < T < 650°C. At this stage, the result of the HT-RT measurement and the XRD analysis do not show any sign of the superconducting property or the $MgB_2$ phase. The superconducting phase, therefore, exists in a microscopic scale without forming a continous path throughout the sample. The second stage of the phase formation process can be classified from 650°C, the appearing temperature of the $MgB_2$ phase in a macroscopic scale, to the temperature, 750°C, at which traces of the magnesium and boron phases disappear completely. This is revealed in the XRD analysis, in consistent with the result from the HT-RT measurement. The optimum stage for the fabrication of the $MgB_2$ happens in the temperature range, 750°C < T < 900°C, in which the Mg and B have been consumed completely for the HT-RT measurement to exhibit minimum resistivity. For T > 900°C, the $MgB_2$ phase is actually suppressed or destroyed due to the escaping of the magnesium vapor. This is indicated by the sudden reduction of the Meissner diamagnetizm shown in Fig. 3 at T = 980°C.

The current investigations on the phase formation process of the polycrystalline $MgB_2$ during fabrication provide revealing and important information for the sample preparation. It is of particular interests that the phase formation temperature range is very narrow for the $MgB_2$ as compared to that for the YBCO high $T_C$ superconductor[27].

**Acknowledgement**

This research was a part project of the Department of Physics of Peking University, and is supported by the Center for Research and Development of Superconductivity in China under contract No. BKBRSF-G19990646 and G1999064602.

**Figure captions:**

Fig. 1. The series of XRD spectra for the MgB$_2$ samples quenched at different temperatures during the sintering process. The bottom one is for the pre-sintered MgB$_2$ embryo box, in which the peaks labeled 1 are for the magnesium, and 2 for the boron. The index in the spectrum of T = 800 °C is for the MgB$_2$ phase.

Fig. 2. The SEM images for the samples at different temperatures of the sintering process, including the pre-sintered raw materials, the ones with the sintering temperature at 600°C, 653°C, 660°C, 750℃, and 980°C. The magnification factor is 50000.

Fig. 3 Temperature dependent magnetization for the Meissner state of the samples quenched at various sintering temperatures. The sintering temperatures for the samples are 450°C, 530°C, 600°C, 620°C, 646°C, 653°C, 660°C, 680°C, 700°C, 750°C, 800°C, 900°C, and 980°C, respectively.

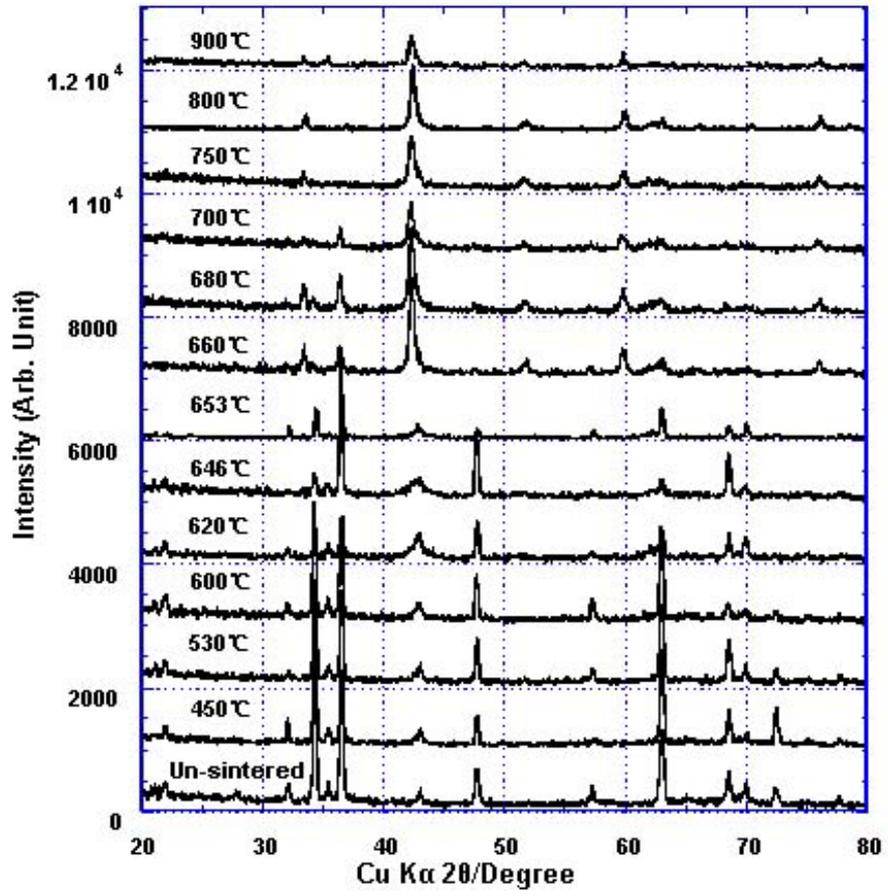

Fig. 1

Q.F. Feng et al.

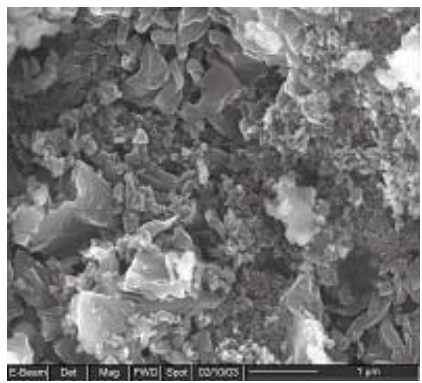
Raw

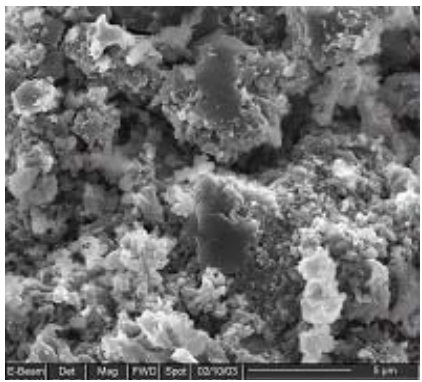
600℃

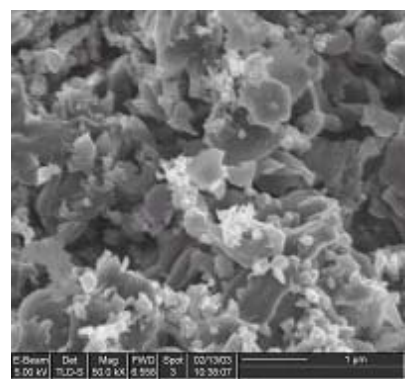
653℃

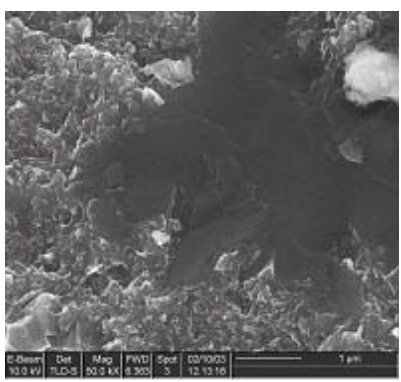
660℃

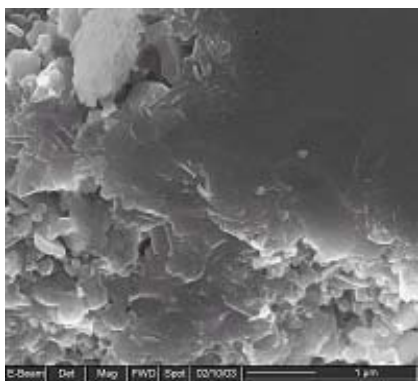
750℃

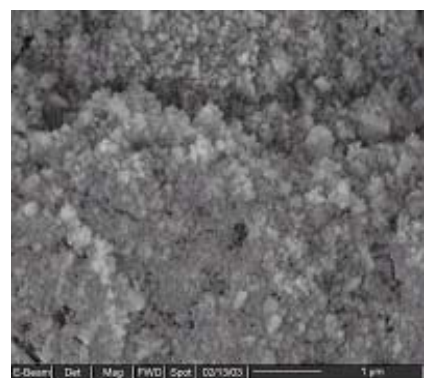
980℃

Fig. 2

Q.F. Feng et al.

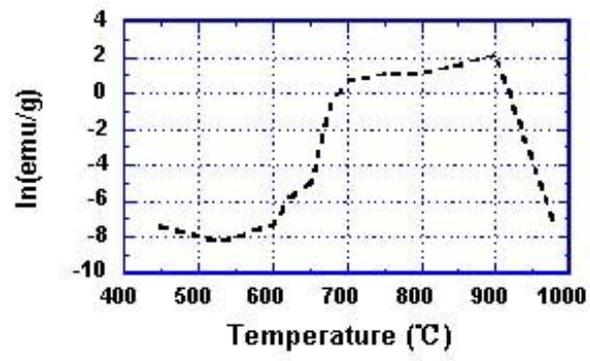

Fig. 3

Q.F. Feng et al.